\documentclass[prd,aps,showpacs,nofootinbib,preprintnumbers, superscriptaddress]{revtex4}
\usepackage{graphicx}
\usepackage{amssymb,amsmath,latexsym}
\usepackage{amsfonts}
\usepackage{bm}
\usepackage{color}
\input{colordvi.tex}

\newcommand{\beqa}{\begin{eqnarray}}
\newcommand{\eeqa}{\end{eqnarray}}

\newcommand{\vp}{\varphi}

\newcommand{\beq}{\begin{equation}}  \newcommand{\eeq}{\end{equation}}
\newcommand{\bef}{\begin{figure}}  \newcommand{\eef}{\end{figure}}
\newcommand{\bec}{\begin{center}}  \newcommand{\eec}{\end{center}}

\newcommand{\tilR}{\widetilde{R}}

\newcommand{\tils}{\widetilde{s}}
\newcommand{\tilg}{\widetilde{g}}
\newcommand{\tilt}{\widetilde{t}}
\newcommand{\tilnabla}{\widetilde{\nabla}}

\newcommand{\tilm}{\widetilde{m}}
\newcommand{\tilu}{\widetilde{u}}
\newcommand{\tilv}{\widetilde{v}}
\newcommand{\tilk}{\widetilde{k}}
\newcommand{\tillambda}{\widetilde{\lambda}}
\newcommand{\tilphi}{\widetilde{\phi}}
\newcommand{\tilG}{\widetilde{G}}
\newcommand{\tilnu}{\widetilde{\nu}}
\newcommand{\tilz}{\widetilde{z}}
\newcommand{\tilH}{\widetilde{H}}
\newcommand{\tild}{\widetilde{d}}
\newcommand{\tilL}{\widetilde{L}}
\newcommand{\tilf}{\widetilde{f}}
\newcommand{\tilrho}{\widetilde{\rho}}
\newcommand{\tilp}{\widetilde{p}}
\newcommand{\tilP}{\widetilde{P}}
\newcommand{\tilS}{\widetilde{S}}
\newcommand{\tilT}{\widetilde{T}}
\newcommand{\tilTheta}{\widetilde{\Theta}}
\newcommand{\tilPsi}{\widetilde{\Psi}}
\newcommand{\tilPhi}{\widetilde{\Phi}}
\newcommand{\tilc}{\widetilde{c}}
\newcommand{\tilGamma}{\widetilde{\Gamma}}

\newcommand{\tilzeta}{\widetilde{\zeta}}

\renewcommand{\(}{\left(} 
\renewcommand{\)}{\right)} 
\renewcommand{\[}{\left[} 
\renewcommand{\]}{\right]} 

\begin{document}
%\draft

\title{
Conformal-Frame (In)dependence of \\
Cosmological Observations in Scalar-Tensor Theory
%or Jordan-Einstein Dictionary
}

\author{Takeshi Chiba}
\affiliation{Department of Physics, College of Humanities and Sciences, Nihon University, Tokyo 156-8550, Japan}
\author{Masahide Yamaguchi}
\affiliation{Department of Physics, 
Tokyo Institute of Technology, Tokyo 152-8551, Japan}%

\date{\today}


\begin{abstract}
We provide the correspondence between the variables in the Jordan frame
and those the Einstein frame in scalar-tensor gravity and consider the
frame-(in)dependence of the cosmological observables. In particular, we
show that the cosmological observables/relations (redshift, luminosity
distance, temperature anisotropies) are frame-independent. We also 
study the frame-dependence of curvature perturbations and find that 
the curvature perturbations are conformal invariant if the perturbation is 
adiabatic and the entropy perturbation between matter and the Brans-Dicke 
scalar is vanishing. The relation among various definitions of 
curvature perturbations in the both frames is also discussed, and 
the condition for the equivalence is clarified. 
\end{abstract}

\maketitle

\section{Introduction}

The scalar-tensor gravity \cite{st} is usually formulated in the
so-called Jordan frame where the metric tensor is minimally coupled to
the matter sector.  On the other hand, when we perform the conformal
transformation of the metric, the action can be reduced to that of the
Einstein gravity, but this time the metric is non-minimally coupled to
the matter sector \cite{dicke}.  So, the question is, "which frame
describes the physics?", about which there has been a long debate.

This problem is attacked recently by Deruelle and Sasaki \cite{ds} (see also \cite{armendariz,cps,zg}) by
comparing the observables in the Einstein gravity and those in the
theory "conformally transformed" from the Einstein gravity. They found
that the Hubble's law holds in both frames and therefore the
observational predictions are exactly the same in both frames although
the physical interpretation (redshift and the evolution of the universe,
for example) in each frame is different.

However, the conformal factor studied there is an arbitrary function of
space-time and is not a dynamical field since there is no scalar
gravitational degrees of freedom in general relativity.  So, it is
unclear to what extent their conclusion holds for the scalar-tensor
gravity where the conformal factor corresponds to the scalar
gravitational degrees of freedom of the theory and is truly a dynamical
field.  This is the problem we are going to solve.  Our motivation is
the same in spirit with \cite{dicke,ds}: Since the conformal
transformation is merely a change of variables or units, physical laws
and any observables should not depend on the particular set of variables
and should be conformal-frame independent.

In the following sections, we calculate in both frames the observables
in classical gravitational theory which include the effective gravitational constant (Sec. II), redshift
(Sec. III A.), the luminosity distance and the Hubble's law (Sec. III
B.), the Sachs-Wolfe effect (Sec.IV A.) and the adiabatic/isocurvature
perturbation (Sec. IV B.) and confirm the frame independence or
dependence. In short, we provide a Jordan-Einstein dictionary.  
In Sec. IV B. we also clarify the relation among the various definitions of 
the curvature perturbations in both frames and 
give the condition of the equivalence. 
Some of our results in Sec. III  and IV overlap with the results in
\cite{armendariz,cps,white}. We extend them to include the effective gravitational 
constant, the Hubble's law and the relation among the various definitions of 
the curvature perturbations. 
Several applications of the conformal
transformation are given in Sec. II C. and Sec. II D. We use the units
of $c=\hbar=1$.

\section{Jordan vs. Einstein: the change of unit}

The action of the scalar-tensor theories of gravity in the so-called Jordan frame is given by 
\beqa
S(\tilg_{\mu\nu},\Phi,\psi)=\frac{1}{2\kappa^2}\int d^4x \sqrt{-\tilg}\left(\Phi \tilR-
\frac{\omega(\Phi)}{\Phi}(\tilnabla \Phi)^2-U(\Phi)\right)+S_m(\tilg_{\mu\nu},\psi),
\label{action:j}
\eeqa
where $\kappa^2=8\pi G_*$ is the bare gravitational constant, $\tilg_{\mu\nu}, \Phi$ are the Jordan-frame metric, the Brans-Dicke scalar field, respectively. 
$S_m$ is the matter action and $\psi$ denotes the matter field.  
The equations of motion derived from the action Eq. (\ref{action:j}) are given by
\beqa
&&\Phi\left(\tilR_{\mu\nu}-\frac12 \tilg_{\mu\nu}\tilR\right)=\kappa^2 \tilT_{\mu\nu}
+\frac{\omega(\Phi)}{\Phi}
\left(\partial_{\mu}\Phi\partial_{\nu}\Phi-\frac12\tilg_{\mu\nu}(\tilnabla\Phi)^2\right)
+\tilnabla_{\mu}\tilnabla_{\nu}\Phi-\tilg_{\mu\nu}\widetilde{\Box}\Phi-\frac12 \tilg_{\mu\nu}U, \label{eom:st:1}\\
&&\widetilde{\Box}\Phi=\frac{1}{2\omega(\Phi)+3}\left(\kappa^2 \tilT-\frac{d\omega(\Phi)}{d\Phi}(\tilnabla\Phi)^2
+\Phi\frac{dU}{d\Phi}-2U\right),\label{eom:st:2}
\eeqa
where the energy-momentum tensor $\tilT_{\mu\nu}$ is defined by 
$\sqrt{-\tilg}\tilT_{\mu\nu}=-2\delta S_m/\delta \tilg^{\mu\nu}$ and 
$\tilT={\tilT^{\mu}}_{\mu}$ is the trace, and the matter energy-momentum conservation is
\beqa
\tilnabla_{\mu}{\tilT^{\mu}}_{\nu}=0.
\label{emc:st}
\eeqa

There is another distinct frame so-called Einstein frame where the gravity action becomes 
that of the Einstein gravity with a minimally coupled scalar field. 
The Einstein frame metric $g_{\mu\nu}$ is related to the Jordan frame metric by 
the conformal transformation
\beqa
\tilg_{\mu\nu}=\frac{1}{\Phi}g_{\mu\nu}\equiv e^{2a}g_{\mu\nu}, 
\label{eom:gmunu}
\eeqa 
and the action can be rewritten as 
\beqa
S(g_{\mu\nu},\vp,\psi)=\int d^4x \sqrt{-g}\left(\frac{R}{2\kappa^2}-\frac12(\nabla\vp)^2-V(\vp)\right)
+S_m(e^{2a(\vp)}g_{\mu\nu},\psi),
\label{action:e}
\eeqa
where $\vp$ and $V(\vp)$ are defined by
\beqa
&&\frac{1}{4\Phi^2}\left(\frac{d\Phi}{d\vp}\right)^2=\left(\frac{da}{d\vp}\right)^2=\frac{\kappa^2}{2(2\omega(\Phi)+3)},
\label{Phi-vp}\\
&&V(\vp)=\frac{U(\Phi)}{2\kappa^2 \Phi^2}.
\label{V-U}
\eeqa
The equations of motion are given by
\beqa
&&R_{\mu\nu}-\frac12 g_{\mu\nu}R=\kappa^2\[ T_{\mu\nu}+
\partial_{\mu}\vp\partial_{\nu}\vp-\frac12 g_{\mu\nu}\left((\nabla\vp)^2+2V(\vp)\right)\], \label{eom:ein:1}\\
&&{\Box}\vp-\frac{d V}{d\vp}=-\frac{d a}{d\vp}T,
\label{eom:ein:2}
\eeqa
where the energy-momentum tensor $T_{\mu\nu}$ is defined by 
$\sqrt{-g}T_{\mu\nu}=-2\delta S_m/\delta g^{\mu\nu}$ and  
$T={T^{\mu}}_{\mu}$ is the trace. 

The energy-momentum tensor in the Einstein frame $T_{\mu\nu}$ is related to that in the Jordan frame $\tilT_{\mu\nu}$ as
\beqa
\tilT_{\mu\nu}=e^{-2a}T_{\mu\nu},
\label{eom:tmunu}
\eeqa
and the matter energy-momentum conservation  in the Einstein frame is given by
\beqa
\nabla_{\mu}{T^{\mu}}_{\nu}=\frac{d a}{d\vp}(\partial_{\nu}\vp) T.
\label{emc:ein}
\eeqa

{We can directly observe that Eqs. (\ref{eom:st:1}),
(\ref{eom:st:2}), and (\ref{emc:st}) are equivalent to
Eqs. (\ref{eom:ein:1}), (\ref{eom:ein:2}), and (\ref{emc:ein}) if we
change variables $\tilg_{\mu\nu}=g_{\mu\nu}/\Phi = e^{2a}g_{\mu\nu}$
given in Eq. (\ref{eom:gmunu}) and use the relations
Eqs. (\ref{Phi-vp}), (\ref{V-U}), and (\ref{eom:tmunu}).}

In the following we consider the motion of particle and photon, and
compare the observed redshift in both frames.

\subsection{Particle motion}
The action of test particle of rest mass $\tilm$ in the Jordan frame is given by
\beqa
S_p=\int \tilm d\tils,
\eeqa
and the motion of the particle (with the 4-velocity $\tilu^{\mu}$) in the Jordan frame 
is given by the familiar geodesic equation:
\beqa
\tilu^{\nu}\tilnabla_{\nu}\tilu^{\mu}=0.
\label{particle:jordan}
\eeqa
Since 
\beqa
d\tils^2=\tilg_{\mu\nu} dx^{\mu}dx^{\nu}=e^{2a(\vp)}ds^2,
\label{ds2}
\eeqa 
the action in the Einstein frame is written as
\beqa
S_p=\int \tilm e^{a(\vp)} ds\equiv \int m(\vp)ds,
\label{mass}
\eeqa
and hence the motion of the particle (with the 4-velocity $u^{\mu}(=e^{a(\vp)}\tilu^{\mu})$) in the Einstein 
frame is given by
\beqa
u^{\nu}\nabla_{\nu}u^{\mu}=-(u^{\mu}u^{\nu}+g^{\mu\nu})\partial_{\nu}a(\vp).
\label{particle:einstein}
\eeqa
Eq. (\ref{ds2}) determines the relation of the proper distance/time between 
 both frames. 
The trajectory in the Einstein frame is not a "straight line".  
However, it is incorrect to state that "the (weak) equivalence principle is violated in the Einstein frame \cite{faraoni}", as is clear from studying the Newtonian 
limit given below. 
The weak equivalence principle is violated  when the 
$\vp$-dependence cannot be eliminated by the conformal transformation.

Note that the spatial velocity of matter (the velocity measured 
in the locally orthonormal  frame) 
$\tilv^i\equiv \tilu^i/\tilu^0$ is independent of the conformal-frame, $\tilv^i=v^i$. 

\subsection{Photon}
As is well-known, the classical electromagnetic action is conformally invariant: 
\beqa
S_{EM}=-\frac14 \int d^4x \sqrt{-\tilg} \tilg^{\mu\rho}\tilg^{\nu\sigma}F_{\mu\nu}F_{\rho\sigma}=
-\frac14 \int d^4x \sqrt{-g} g^{\mu\rho}g^{\nu\sigma}F_{\mu\nu}F_{\rho\sigma},
\eeqa
where $F_{\mu\nu}=\partial_{\mu}A_{\nu}-\partial_{\nu}A_{\mu}$, so 
the photon (with the tangent vector $\tilk^{\mu}(k^{\mu})$ in the Jordan (Einstein) frame) 
follows the null geodesics in both frames:
\beqa
\tilk^{\nu}\tilnabla_{\nu}\tilk^{\mu}=0=k^{\nu}\nabla_{\nu}k^{\mu} .
\eeqa

\subsection{Newtonian limit and the Gravitational constant}

Let us consider the Newtonian limit (weak gravity and slow motion and
the weak material stresses) to determine the effective gravitational
constant. We set $U(\Phi)=0$ for simplicity.  In the Jordan frame, the
geodesic equation Eq. (\ref{particle:jordan}) yields the Newtonian
equation of motion if
\beqa
\tilg_{00}\simeq -(1+2\tilphi),~~~~~\tilg_{ij}\simeq \delta_{ij},
\eeqa
where $\tilphi$ is the Newtonian gravitational potential produced by the rest mass density $\tilrho$ as
\beqa
\nabla^2\tilphi=4\pi\tilG\tilrho,
\label{poisson:jordan}
\eeqa
where $\tilG$ is the effective gravitational constant determined later. 

Adopting the post-Newtonian bookkeeping rule \cite{will}, we assign the Newtonian gravitational potential as O(2) quantity. Similarly, $v^2\sim \tilp/\tilrho\sim$O(2) and $|\partial/\partial t|/|\partial/\partial x|\sim$ O(1). 
The Newtonian limit (up to O(2)) of Eq. (\ref{eom:st:2}) becomes
\beqa
\nabla^2\Phi=-\frac{\kappa^2\tilrho}{2\omega(\Phi)+3},
\label{scalar:newton:jordan}
\eeqa
and the solution up to O(2) can be written as
\beqa
\Phi=\Phi_0+\frac{\kappa^2}{4\pi(2\omega_0+3)}\int d^3x'\frac{\tilrho({\bf x'})}{
|{\bf x}-{\bf x'}|},
\eeqa
where $\Phi_0$ denotes the value of $\Phi$ at spatial infinity and we have already expanded $\omega(\Phi)$ and replaced $\omega(\Phi)$ with  its asymptotic 
value $\omega_0\equiv\omega(\Phi_0)$. 
Then, from Eq. (\ref{eom:st:1}) using Eq. (\ref{scalar:newton:jordan}) we obtain
\beqa
\Phi_0\nabla^2\tilphi=\frac{\kappa^2}{2}\frac{2\omega_0+4}{2\omega_0+3}\tilrho.
\eeqa
Comparing this with Eq. (\ref{poisson:jordan}), we can determine the effective gravitational constant \cite{de} \footnote{Our $\vp$ is related to $\vp_{\rm DEF}$ used in \cite{de} via $\vp=\sqrt{2}\vp_{\rm DEF}/\kappa$.}
\beqa
\tilG=\frac{\kappa^2}{8\pi }\frac{2\omega_0+4}{2\omega_0+3}\frac{1}{\Phi_0}
=G_*\left[e^{2a(\vp)}\left(1+\frac{2}{\kappa^2}\left(\frac{da}{d\vp}\right)^2\right)\right]_{\vp_0},
\label{G:jordan}
\eeqa
where $\vp_0$ is the asymptotic value of $\vp$ corresponding to $\Phi_0$ and we have used Eq. (\ref{Phi-vp}) and $\kappa^2=8\pi G_*$. 
This $\tilG$ is the quantity which is 
measured by the Cavendish-type experiment. Since $\tilG$ involves $\vp$, the mass 
of a self-gravitating body depends on $\vp$, and hence the strong 
equivalence principle is violated in the scalar-tensor theory. 

Next, we consider the Newtonian limit in the Einstein frame. 
For the metric in the Newtonian limit, $g_{00}\simeq -(1+2\phi), g_{ij}\simeq \delta_{ij}$, 
the Newtonian limit of Eq. (\ref{eom:ein:1}) and Eq. (\ref{eom:ein:2}) become
\beqa
&&\nabla^2\phi=\frac{\kappa^2}{2}\rho,\\
&&\nabla^2\vp=\frac{da}{d\vp}\rho.
\eeqa
Similar to $\Phi$, the solution of $\vp$ up to O(2) is given by
\beqa
\vp=\vp_0-\frac{1}{4\pi}\frac{da}{d\vp}\Big{|}_0\int d^3x'\frac{\rho({\bf x'})}{
|{\bf x}-{\bf x'}|}.
\eeqa
Then the particle equation of motion Eq.(\ref{particle:einstein}) in the Newtonian limit becomes
\beqa
\frac{d^2{\bf x}}{dt^2}&=&-\nabla\phi-\frac{da}{d\vp}\nabla\vp \nonumber\\
&=&-\frac{\kappa^2}{8\pi}\left(1+\frac{2}{\kappa^2}\left(\frac{da}{d\vp}\Big{|}_0\right)^2\right)\int d^3x'\frac{({\bf x}-{\bf x'})\rho({\bf x'})}{|{\bf x}-{\bf x'}|^3}.
\eeqa
The coefficient in front of the integral is to be regarded as the effective gravitational constant $G$ in the Einstein frame:  
\beqa
G=G_*\left[1+\frac{2}{\kappa^2}\left(\frac{da}{d\vp}\right)^2\right]_{\vp_0}.
\label{G:einstein}
\eeqa
$\tilG$ and $G$ are related as $G=e^{-2a(\vp_0)}\tilG$ in accord with the fact that 
the gravitational constant has the dimension of $({\rm mass})^{-2}$ and the unit of mass changes by changing the frame  according to Eq. (\ref{mass}).  However, the transformation may be regarded as "non-local" 
in the sense that the conformal transformation refers to the value of $\vp$ at spatial infinity.

\section{Cosmology in Scalar-Tensor Theory}

\subsection{Redshift}

Now we consider the observed redshift in each frame. We calculate the redshift in the FRW spacetime. 
The line element in the Jordan frame is given by
\beqa
d\tils^2=-d\tilt^2+\tilR(\tilt)^2(d\chi^2+\sin^2_K\chi d\Omega^2)=\tilR(\eta)^2(-d\eta^2+d\chi^2+\sin^2_K\chi d\Omega^2),
\eeqa
where $\tilt$ is the cosmic time and $\tilR$ is the scale factor and $\eta$ is the conformal time. $\sin_K\chi=\sin\chi(\sinh\chi)$ for a closed $(K=1)$ (an open $(K=-1)$ ) universe and $\sin_K\chi=\chi$ for a flat universe $(K=0)$. 
Since  $d\tils^2=\tilg_{\mu\nu}dx^{\mu}dx^{\nu}=e^{2a(\vp)}ds^2$, the corresponding variables (without tilde) in the Einstein frame are given by
\beqa
d\tilt=e^{a(t)}dt,~~~~~\tilR(\tilt)=e^{a(t)}R(t) .
\label{tau:scale}
\eeqa

First, we calculate the observed redshift in the Jordan frame. For the null geodesic with 
the tangent vector $\tilk^{\mu}$, the frequency $\tilnu$ 
measured by the observer with 4-velocity $\tilu^{\mu}$ is given by $\tilnu=-\tilk^{\mu}\tilu_{\mu}$. Then  
the observed redshift is given by
\beqa
1+\tilz=\frac{\tilnu_S}{\tilnu_O},
\eeqa
where the subscript $S(O)$ denotes the quantity measured at the source (observer's) position. 
The derivative of $\tilnu$ with respect to the affine parameter $\tillambda$ is given by
\beqa
\frac{d\tilnu}{d\tillambda}=-\tilk^{\mu}\tilk^{\nu}\tilnabla_{\mu}\tilu_{\nu},
\eeqa
Since $\tilu^{\mu}$ obeys the geodesics in the FRW universe, $\tilnabla_{\mu}\tilu_{\nu}=\tilH(\tilg_{\mu\nu}+\tilu_{\mu}\tilu_{\nu})$, where 
$\tilH=\tilR^{-1}(d\tilR/d\tilt)$ is the Hubble parameter in the Jordan frame. So we have \cite{wald}
\beqa
\frac{d\tilnu}{d\tillambda}=-\tilnu^2\tilH,
\eeqa
or
\beqa
\frac{d\tilnu}{d\tilt}=-\tilnu\tilH.
\eeqa
Hence, we obtain the usual redshift law: $1+\tilz\propto \tilR^{-1}$. Therefore, 
the observed redshift is given by
\beqa
1+\tilz=\frac{\tilnu_S}{\tilnu_O}=\frac{\tilR_O}{\tilR_S}.
\label{redshift:jordan}
\eeqa

Next, we calculate the redshift in the Einstein frame. Now that the observer's 4-velocity $u^{\mu}$ does not follow geodesics, and $\nabla_{\mu}u_{\nu}$ is given by 
$\nabla_{\mu}u_{\nu}=H(g_{\mu\nu}+u_{\mu}u_{\nu})-u_{\mu}(u^{\rho}\nabla_{\rho}u_{\nu})$, where 
$H=R^{-1}(dR/dt)$. \footnote{Note that $H$ is ${\it not}$ the measurable Hubble parameter 
in the Einstein frame. See the discussion below. }
 %is the Hubble parameter in the Einstein frame. 
So the derivative of $\nu$ with respect to the affine parameter $\lambda$ becomes
\beqa
\frac{d\nu}{d\lambda}&=&-k^{\mu}k^{\nu}\nabla_{\mu}u_{\nu},\nonumber\\
&=&-\nu^2H-\nu k^{\mu}u^{\rho}\nabla_{\rho}u_{\mu}.
\eeqa
Using Eq. (\ref{particle:einstein}), we find that the last term vanishes. Hence again $\nu\propto R^{-1}$ and 
\beqa
\frac{\nu_S}{\nu_O}=\frac{R_O}{R_S}.
\label{redshift-o:einstein}
\eeqa
So, considering Eq. (\ref{tau:scale}), one may naively conclude that the observed is conformal-frame dependent. 
However, it is incorrect, since the observation is (and should be) made in the observer's reference frame where 
the unit is in general different from that of the source frame. 
For example, from Eq. (\ref{tau:scale}), for the same transition frequency $\tilnu$,  the frequency measured 
in the unit of one frame is $\nu_1=e^a|_1\tilnu$ and the frequency measured in the unit of another frame is 
$\nu_2=e^a|_2\tilnu$, and the relation between them is \cite{ds}
\beqa
\frac{\nu_1}{\nu_2}=\frac{e^a|_1}{e^a|_2}.
\label{relation}
\eeqa
The observed redshift is, more precisely, defined by 
the ratio of the source frequency in unit of the observer's frame to the observed frequency. 
It is given by, taking into account the relation Eq. (\ref{relation}), 
\beqa
1+z=\frac{e^a|_O}{e^a|_S}\frac{\nu_S}{\nu_O}=\frac{e^a|_O}{e^a|_S}\frac{R_O}{R_S}=\frac{\tilR_O}{\tilR_S}=1+\tilz ,
\label{redshift:einstein}
\eeqa 
where we have used Eq. (\ref{redshift-o:einstein}) in the second equality. 
Therefore, the observed redshift is conformal-frame independent \cite{armendariz,cps}. 

\subsection{Distances}

\subsubsection{Luminosity distance and Hubble's Law}

The luminosity distance in the Jordan frame $\tild_L$ is defined in terms of the source's (absolute) luminosity $\tilL_S$ and the observed flux $\tilf_O$ as
$\tild_L=\sqrt{\tilL_S/4\pi \tilf_O}$. Since $\tilL_S$ is redshifted to become the  luminosity in the observer's frame $\tilL_O=\tilL_S(\tilR_S/\tilR_O)^2$ and 
$\tilf_O=\tilL_O/4\pi \tilR_O^2\sin^2_K\chi$, $\tild_L$ is written in the usual form
\beqa
\tild_L=\tilR_O(1+\tilz)\sin_K\chi,
\label{dL}
\eeqa
where $\chi$ is the comoving distance to the source and is given by
\beqa
\chi=\int^{\tilt_O}_{\tilt_S}\frac{d\tilt}{\tilR}=\int^{\tilz}_{0}
\frac{d\tilz}{\tilR_O\tilH(\tilz)}.
\label{chi}
\eeqa
Taking the limit $\tilz\rightarrow 0$, 
we obtain the Hubble's law: 
\beqa
\tild_L=\frac{\tilz}{\tilH_0},
\eeqa
where $\tilH_0=\tilH(0)$ is the present Hubble parameter. 

Now we calculate the luminosity distance $d_L$ in the Einstein frame. 
Remembering the discussion about the redshift, one should consistently use the unit 
of the observer's frame to define the luminosity distance. Since the luminosity has the unit of mass divided by time, the source's luminosity $L_S$ measured by the observer's unit is $L_S(e^a|_O/e^a|_S)^2$ and from Eq. 
(\ref{redshift-o:einstein}), $L_S=L_O(R_O/R_S)^2$.  Thus
\beqa
d_L=\sqrt{\frac{L_S}{4\pi f_O}\left(\frac{e^a|_O}{e^a|_S}\right)^2}=
R_O\frac{R_O}{R_S}\frac{e^a|_O}{e^a|_S}\sin_K\chi=R_O(1+z)\sin_K\chi,
\eeqa
where $\chi$ is the same as Eq. (\ref{chi}) since the null geodesics 
is conformally invariant. 
Note that since $R=e^{-a}\tilR$,  
$d_L$ differs from $\tild_L$: $d_L=e^{-a}|_O\tild_L$. 
But this is legitimate because the unit is different between the frames and 
this is precisely the correspondence of length (dimensionful quantity) in the Einstein frame to that in the Jordan frame. 

Again taking $z\rightarrow 0$, we find
\beqa
d_L=\frac{z}{e^{a}|_O\tilH_0}\equiv \frac{z}{H_{E0}}.
\eeqa
{}From Eq. (\ref{tau:scale}), the Einstein-frame Hubble parameter $H_{E0}$ 
is given by
\beqa
H_{E0}=e^a|_O\tilH_0=\frac{1}{\tilR}\frac{d\tilR}{dt}\Big{|}_0=
H_0+\frac{da}{dt}\Big{|}_0.
\label{hubble:einstein}
\eeqa
$H_{E0}$ differs from $\tilH_0$ by $e^a|_O$ 
in accord with the fact that the Hubble parameter has the dimension of the inverse of time.

Note that since $e^{a}|_O$ is merely a constant, it is absorbed into the definition of the 
absolute magnitude, and hence the magnitude-redshift relation is 
conformal-frame independent. 

The correspondence between the variables in the Jordan frame and those in the Einstein frame is given in Table \ref{table1}. 

\subsubsection{Angular Diameter Distance and the Reciprocity Relation}

For an object with the proper diameter $\widetilde{D}$ and 
the angular diameter $\theta$, the angular diameter distance $\tild_A$ in the Jordan frame is defined by
\beqa
\tild_A\equiv \frac{\widetilde{D}}{\theta}=\tilR_S\sin_K\chi=\frac{1}{1+\tilz}\tilR_O\sin_K\chi,
\eeqa
where we have used Eq. (\ref{redshift:jordan}). 
The angular diameter distance is related to the luminosity distance through 
the so-called reciprocity relation \cite{ellis}, $\tild_L=(1+\tilz)^2\tild_A$,  which holds as long as the photon energy-momentum conservation holds.

The discussion similar in the luminosity distance gives the proper diameter 
in the observer's unit in the Einstein frame as
\beqa
D=\frac{e^a|_S}{e^a|_O}R_S\sin_K\chi\theta.
\eeqa
The angular diameter distance in the Einstein frame is given by
\beqa
d_A=\frac{e^a|_S}{e^a|_O}R_S\sin_K\chi=\frac{1}{1+\tilz}R_O\sin_K\chi=
e^{-a}|_O\tild_A.
\eeqa
Since the photon energy-momentum conservation also holds in the Einstein frame, 
the reciprocity relation also holds: $d_L=(1+z)^2d_A$.

%%%%%%%%%%%%%%%%%%%%
\begin{table}
\begin{center}
\begin{tabular*}{0.7\textwidth}{@{\extracolsep{\fill}}lcc}
\hline
\hline
quantity & in Jordan frame & in Einstein frame \\
\hline
metric &  $\tilg_{\mu\nu}$ & $g_{\mu\nu}=e^{-2a}\tilg_{\mu\nu}$ \\
time/length &  $d\tils$  &  $ds=e^{-a}d\tils$ \\
scale factor &  $\tilR$ &  $R=e^{-a}\tilR$ \\
(particle) mass & $\tilm$ &  $m=\tilm e^a$ \\
effective gravitational constant& $\tilG$ & $G=e^{-2a(\vp_0)}\tilG$\\
energy density & $\tilrho$ &  $\rho=e^{4a}\tilrho$ \\
redshift  &  $\tilz$  &  $z=\tilz$ \\
Hubble parameter  &  $\tilH$ & $H_E=H+da/dt=e^a\tilH$\\
curvature perturbation & $\tilzeta$ & $\zeta=\tilzeta$ \\
  &  &   (if $\delta\rho/(d\rho/dt)=\delta a/(da/dt)$)\\

\hline
\hline
\end{tabular*}
\caption{Jordan-Einstein dictionary. The scaling relation of dimensionful quantity can be found 
by the change of unit of length, time,  mass given in the second to the fourth three columns. $\tilG$ is given by Eq. (\ref{G:jordan}). 
\label{table1}}
\end{center}
\end{table}
%%%%%%%%%%%%%%%%%%%%%%%%%%%%%%%%%%%

\subsection{Horizon problem and inflation}

As an exercise, let us consider the horizon problem and the condition
for inflation. The almost isotropy of the cosmic microwave background
indicates that the Universe was already smooth at the decoupling time
$\tilt_{dec}$. The necessary condition for solving the horizon problem
is that the distance light can travel between two times (say, from
$\tilt_i$ and $\tilt_f(<\tilt_{dec})$) is larger than the present Hubble
distance. In the comoving coordinate, the condition is given by
\cite{htw}
\beqa
\int_{\tilt_i}^{\tilt_f}\frac{d\tilt}{\tilR}>\frac{1}{\tilR_0\tilH_0}. 
\label{eq:Jorhor}
\eeqa
We rewrite the condition in terms of the Einstein-frame variables.
Since from Eq. (\ref{tau:scale}) and Eq. (\ref{hubble:einstein}),
$\tilR_0\tilH_0=R_0H_{E0}$, the condition is conformal-frame independent
\beqa
\int_{t_i}^{t_f}\frac{dt}{R}>\frac{1}{R_0H_{E0}}.
\eeqa
In the comoving coordinate, the wavenumber of any mode remains constant with time. 
In the standard cosmology, since the comoving Hubble distance $(\tilR\tilH)^{-1}$ increases with time, these modes must have all been outside the horizon before the decoupling. Then the necessary requirement to solve the horizon problem is that  $(\tilR\tilH)^{-1}$  must have decreased sometime in the past. 
The requirement is written as 
\beqa
\frac{d}{d\tilt}\left(\frac{1}{\tilR\tilH}\right)=
-\left(\frac{d\tilR}{d\tilt}\right)^{-2}\frac{d^2\tilR}{d\tilt^2}<0
\eeqa
in the past, or $d^2\tilR/d\tilt^2>0$, which is nothing but the condition of  inflation. Alternatively, we may consider the condition in the contracting universe ($\tilH<0$): $d(-(\tilR\tilH)^{-1})/d\tilt<0$, 
or $d^2\tilR/d\tilt^2<0$.  

In the Einstein frame, the comoving Hubble distance is $(\tilR\tilH)^{-1}=(RH_E)^{-1}$, where $H_E=H+da/dt$ (see Eq. (\ref{hubble:einstein})). Therefore, the requirement that the comoving Hubble distance decreases with time is then
\beqa
\frac{d}{dt}\left(\frac{1}{RH_E}\right)<0,
\eeqa
or $d^2R/dt^2+(dR/dt)(da/dt)+Rd^2a/dt^2>0$, and the opposite inequality for 
the contracting universe. Therefore, it is possible in principle to solve 
the horizon problem without accelerating expansion in the Einstein frame. 
An extreme example as such is the pre-big-bang scenario \cite{veneziano} 
which corresponds to $\omega=-1$ Brans-Dicke gravity. 
The solution in a flat universe  is given by
\beqa
\tilR=(-\tilt)^{-1/\sqrt{3}},~~~~~\tilH=\frac{1}{\sqrt{3}(-\tilt)},~~~~~\Phi=(-\tilt)^{1+\sqrt{3}},
\eeqa
for $\tilt<0$, whereas  the universe contracts in the Einstein frame
\beqa
R=(-t)^{1/3},~~~~~H=-\frac{1}{3(-t)}.
\eeqa
Although it is possible to solve the horizon problem, the pre-big-bang requires 
a fine-tuning of the initial conditions in order to account for the 
 the flatness and homogeneity of our part of the universe \cite{turner}.

\subsection{Apparent Equation of State}

As a second application, let us consider the equation of state of dark energy if 
$\Phi$ (or $\vp$) plays the role of dark energy.   
The basic equations in FRW universe models are given in Appendix. 

An interaction between dark matter and dark energy can result 
in an effective dark energy equation of state of $w<-1$ \cite{das}. 
In the Einstein frame, $\vp$ couples to (dark) matter, so the scalar-tensor gravity in the Einstein frame provides such an interaction.\footnote{Note again that this does not immediately imply that the weak 
equivalence principle is violated. See the discussion of the Newtonian limit in Sec.II C.} 
To see how $w<-1$ happens, consider the situation where an observer (erroneously) interprets the observational data using the Einstein gravity with a minimal coupling to matter (that is $S_m(g_{\mu\nu},\psi)$ in Eq. (\ref{action:e}) instead of $S_m(e^{2a}g_{\mu\nu},\psi)$).\footnote{In the Jordan frame language, "an observer interprets the observational data using the Einstein gravity with the metric $g_{\mu\nu}$ instead of $\tilg_{\mu\nu}$", and the rest is the same.} 
Then, such an observer regards $H$ as the "correct" Hubble parameter and 
assumes that the dark matter is non-interacting and its energy density 
redshifts as $R^{-3}$. So, instead of Eq. (\ref{friedmann:e})  
the observer assumes the Friedmann equation (in a flat universe consisting of dark matter $\rho_m$ and dark energy $\rho_{\vp}$) as
\beqa
3H^2=\kappa^2\left(\rho_{m0}\left(\frac{R_0}{R}\right)^3+\rho_{\vp}^{\rm eff}\right),
\eeqa 
and by equating this with Eq. (\ref{friedmann:e}), the he/she  would infer an effective dark energy density with 
\beqa
\rho^{\rm eff}_{\vp}=\rho_m-\rho_{m0}\left(\frac{R_0}{R}\right)^3+\rho_{\vp}.
\label{rho:eff}
\eeqa
Assuming that dark energy is also non-interacting, the observer determines the equation of state of dark energy $w_{\rm eff}$ by
\beqa
\frac{d\rho_{\vp}^{\rm eff}}{dt}=-3H(1+w_{\rm eff})\rho^{\rm eff}_{\vp}.
\eeqa
{}From the time derivative of Eq. (\ref{rho:eff}) and noting that 
\beqa
&&\frac{d\rho_m}{dt}=-3H\rho_m+\frac{da}{dt}\rho_m,\label{rho_m}\\
&&\frac{d\rho_{\vp}}{dt}=-3H\left(\frac{d\vp}{dt}\right)^2-\frac{da}{dt}\rho_m
\equiv-3H(1+w_{\vp})\rho_{\vp}-\frac{da}{dt}\rho_m,
\eeqa
$w_{\rm eff}$ is given by
\beqa
w_{\rm eff}=\frac{w_{\vp}\rho_{\vp}}{\rho_m-\rho_{m0}\left(\frac{R_0}{R}\right)^3+\rho_{\vp}}.
\eeqa
Therefore, $w_{\rm eff}<w_{\vp}$ is possible if $\rho_m<\rho_{m0}\left(\frac{R_0}{R}\right)^3$. 
Note that from Eq. (\ref{tilrho_m}) (or Eq. (\ref{rho_m})), $\rho_m$ redshifts as 
\beqa
\rho_m=e^{4a}\tilrho_m=\rho_{m0}\frac{e^{4a}}{e^{4a_0}}\left(\frac{R_0e^{a_0}}{Re^a}\right)^3=\rho_{m0}\frac{e^a}{e^{a_0}}\left(\frac{R_0}{R}\right)^3.
\eeqa
Hence, if $a(\vp)$ is an increasing function of time, then $w_{\rm eff}<w_{\vp}$ 
and $w_{\rm efff}$ can be $w_{\rm eff}<-1$ if $\vp$ slow-rolls so that $w_{\vp}\simeq -1$ \cite{das}. 
This can be easily realized with $a(\vp)$ being an increasing function and  $V(\vp)$ being a decreasing function of $\vp$. 

\section{Perturbed observables}

\subsection{Sachs-Wolfe effect}

Next we consider perturbed observable quantities. First we consider 
the temperature anisotropies of the cosmic background radiation.  
First of all, we note that the distribution function $f$ is conformally invariant because the number of photons is invariant: $dN=fd^3xd^3p$ and that 
the collisionless Boltzmann equation is conformal-frame invariant \cite{cps}:
\beqa
\frac{Df}{D\tilt}\equiv \frac{\partial f}{\partial \tilt}+{\bf v} \cdot
\frac{\partial f}{\partial {\bf x}}+\frac{d{\bf p}}{d\tilt}\cdot\frac{\partial f}{\partial {\bf p}}=\frac{1}{e^a}\frac{Df}{Dt}=0.
\label{Boltzmann}
\eeqa 
Therefore, the temperature anisotropies should be invariant under the conformal transformation. In the following we shall verify the invariance in 
the analysis of linear perturbations. 

In the Jordan frame, the fractional temperature fluctuation for a blackbody, $\tilTheta=\delta \tilT/\tilT$, 
is defined in terms of the photon distribution function by
\beqa
f({\bf x},{\bf \tilp},\tilt)=\frac{1}{\exp\(\frac{\tilp}{\tilT(\tilt)
(1+\tilTheta)}\)-1},
\eeqa
where $\tilp$ is the photon momentum. 
In the longitudinal gauge, the metric can be written as 
\beqa
d\tils^2=-(1+2\tilPsi)d\tilt^2+\tilR^2(1+2\tilPhi)d{\bf x}^2. \label{eq:metricpert}
\eeqa
The collisionless Boltzmann equation  is then written as \cite{ks1}
\beqa
\frac{d}{d\eta}(\tilTheta+\tilPsi)=\frac{\partial \tilPsi}{\partial\eta}-
\frac{\partial \tilPhi}{\partial\eta},
\eeqa
where $d\eta=d\tilt/\tilR$ is the conformal time. 

We now consider the corresponding relation in the Einstein frame. 
Since the temperature has the dimension of mass, $\tilT$ is related to the temperature in the Einstein frame $T$ as $\tilT=e^{-a}T$. So
\beqa 
\tilTheta=\frac{\delta T}{T}-\delta a=\Theta-\delta a,
\eeqa
where $\delta a=a(t,{\bf x})-a(t)$. 
On the other hand, considering $d\tilt=e^{a(t)}dt$ and $\tilR(\tilt)=e^{a(t)}R(t)$,
 $\tilPsi$ and $\tilPhi$ are related to $\Psi$ and $\Phi$ (not to be confused with the Brans-Dicke scalar field) as \cite{cy1}
\beqa
&&\tilPsi=\Psi+\delta a,\\
&&\tilPhi=\Phi+\delta a.
\label{Phi}
\eeqa
Therefore, as expected, the Boltzmann equation is conformal-frame independent. 
\beqa
\frac{d}{d\eta}(\Theta+\Psi)=\frac{\partial \Psi}{\partial\eta}-
\frac{\partial \Phi}{\partial\eta}.
\eeqa
and hence the temperature anisotropy is conformal-frame independent. 
One may also verify that the collisional term is  
also conformal-frame independent because the cross section scales as $e^{2a}$ and 
the momentum integral scales as $e^{-3a}$ and these in total 
give a factor $e^{-a}$ in accord with Eq. (\ref{Boltzmann}) \cite{cps}. 

\subsection{Adiabatic and isocurvature perturbations}

We then consider the adiabatic perturbation and the entropy
perturbation, and ask whether these perturbations are conformal-frame
independent.

The pressure perturbation can be split into adiabatic and entropic parts
as \cite{ks,wmll}
\beqa
\delta \tilp=\tilc_s^2\delta\tilrho+\frac{d\tilp}{d\tilt} \tilGamma,
\label{entropy}
\eeqa
where $\tilc_s^2\equiv (d\tilp/d\tilt)/(d\tilrho/d\tilt)$ and the first
term is the adiabatic part and the second term is entropy (the
non-adiabatic) part.

\subsubsection{Adiabaticity and Curvature Perturbations}

First, we rewrite the gauge invariant total entropy perturbation $\tilGamma$
defined by Eq. (\ref{entropy}), which represents the displacement between hypersurfaces of uniform pressure and uniform energy density, 
in terms of the variables in the Einstein frame:
\beqa
\tilGamma&=&\frac{\delta\tilp}{d\tilp/d\tilt}-\frac{\delta\tilrho}{d\tilrho/d\tilt}\nonumber\\
&=&\frac{e^a(d\rho/dt)}{((dp/dt)-4(da/dt)p)((d\rho/dt)-4(da/dt)\rho)}\left(
{\delta p}-c_s^2{\delta\rho}-\frac{4(da/dt)\rho}{d\rho/dt}({\delta p}-{w}{\delta\rho})-{4\rho
\delta a}(w-c_s^2)
\right),\nonumber\\
&=&\frac{e^a(d\rho/dt)}{((dp/dt)-4(da/dt)p)((d\rho/dt)-4(da/dt)\rho)}\left(
(dp/dt)\Gamma\left(1-4\frac{(da/dt)\rho}{d\rho/dt}\right)+4(da/dt)\rho(w-c_s^2)
\left(\frac{\delta\rho}{d\rho/dt}-\frac{\delta a}{da/dt}
\right)\right),
\eeqa
where $w\equiv\tilp/\tilrho=p/\rho$ and $c_s^2\equiv (dp/dt)/(d\rho/dt)$. 
Therefore, when the entropy perturbation is vanishing in
the Jordan frame $\tilGamma=0$, then $\Gamma=0$ holds also in the Einstein
frame if 
\beqa
w=c_s^2~~~~{\rm or} ~~~~\frac{\delta\rho}{d\rho/dt}=\frac{\delta a}{da/dt}=\frac{\delta\varphi}{d\varphi/dt}.
\label{gammazero}
\eeqa
The former condition corresponds to $w=$ const. 
For example, even if we assume the barotropic equation of 
state $\tilp=\tilp(\tilrho)$ in the Jordan frame,  the equation of state in 
the Einstein frame is then $p=e^{-4a}\tilp(e^{4a}\rho)=p(\rho,\vp)$ and 
the pressure becomes a function of the density and the scalar field $\vp$. 

On the other hand, the latter condition corresponds to  the situation where 
the entropy perturbation between matter and $\vp$ is vanishing. 
In fact, it can be shown that the curvature perturbation is conformal-frame invariant  if this condition holds and moreover it is constant 
if  the adiabatic condition ($\Gamma=0$) holds \cite{wmll}.\footnote{In the absence of matter (i.e. only the Brans-Dicke scalar is present), it is shown that the curvature perturbation 
on comoving hypersurfaces is invariant under the conformal transformation \cite{ms,fk,cy1}.}
The (gauge invariant) curvature perturbation on uniform-density 
hypersurfaces (in the Jordan frame) is defined by
\beqa
\tilzeta\equiv \tilPhi-\frac{\delta\tilrho}{d\tilrho/d\tilt}\tilH,
\eeqa
which is shown to be constant on large scales if the pressure 
perturbation is adiabatic ($\tilGamma=0$)
because the energy-momentum tensor in the Jordan frame is conserved \cite{wmll}: $\tilnabla_{\mu}{\tilT^{\mu}}_{\nu}=0$.
Using the relation Eq. (\ref{Phi}) and Eq. (\ref{tilrho_m}), $\tilzeta$ can be rewritten as (again not to be confused $\Phi$ with the Brans-Dicke scalar field)
\beqa
\tilzeta=\Phi+\delta a+\frac{1}{3(1+w)}\left(\frac{\delta\rho}{\rho}-4\delta a\right).
\eeqa
Using Eq. (\ref{em-conserv:e}), the curvature perturbation on 
comoving hypersurfaces in the Einstein frame can be written as
\beqa
\zeta=\Phi-\frac{\delta\rho}{d\rho/dt}H=\Phi+\frac{\delta\rho }{3H(1+w)\rho-(1-3w)(da/dt)\rho}H.
\eeqa
Note that generally 
$\zeta$ is not constant on large scales even if $\Gamma=0$ because 
the energy-momentum tensor in the Einstein frame is no 
longer conserved in general: $\nabla_{\mu}{T^{\mu}}_{\nu}=\frac{d
a}{d\vp}(\partial_{\nu}\vp) T$  (Eq. (\ref{emc:ein})).  
However, it is found that the two curvature perturbations $\tilzeta$ and $\zeta$ 
are equal if and only if 
\beqa
(3w-1) \left( \frac{\delta\rho}{d\rho/dt}- \frac{\delta a}{da/dt} \right)=0,
\label{eq:adicond}
\eeqa 
so that the entropy perturbation between matter and $\vp$ is vanishing,
or $w=1/3$.  Moreover they are constant if $\tilGamma=0$ and hence
$\Gamma=0$ (from Eq.~\ref{gammazero}) so that the adiabatic condition
holds. For perturbations obeying $\delta\rho/(d\rho/dt)=\delta
a/(da/dt)$, or, when $w=1/3$, the adiabatic perturbation in one frame
remains adiabatic in another frame, and the curvature perturbations in
two frames are equal and constant. One may verify the constancy of
$\zeta$ by the direct calculation of the evolution of $\zeta$ from 
Eq. (\ref{emc:ein}) as,
\beqa
  \frac{d\zeta}{dt} &=& - \frac{H+da/dt}{\rho+p} \frac{dp}{dt}\Gamma
    - \frac{1}{3R^2}\nabla^2 v \nonumber \\
  && + \frac{1}{3H}\frac{1-3w}{1+w}\frac{(da/dt)^2}{\delta a}
  \left[ H\left(\frac{d\delta a/dt}{da/dt} -
  \frac{d}{dt}\left(\frac{\delta\rho}{d\rho/dt} \right) \right) 
         \left(\frac{\delta a}{da/dt} - \frac{\delta\rho}{d\rho/dt}
         \right)
       - \frac{\delta\rho}{d\rho/dt} \frac{d}{dt}
          \left(\frac{\delta a}{da/dt} - \frac{\delta\rho}{d\rho/dt} \right)
   \right],
\eeqa
where $v$ is a velocity scalar perturbation. Thus, as long as the
condition (\ref{eq:adicond}) is satisfied and $\Gamma=0$, $\zeta$ is
conserved on superhorizon scales.  

In sum, the notion of adiabaticity
is {\it not} conformal-frame independent in general, though it is
gauge-invariant \cite{white}.%\footnote{However,
%it should be noted that we refer to the correspondence between
%the curvature perturbations on uniform-(matter) density hypersurfaces. 
%The energy density is determined by the corresponding energy-momentum 
%tensor defined by the matter action. We do not refer to the 
%"effective" energy-momentum tensor $S_{\mu\nu}$ 
%defined by rewriting the field equation in the form of the Einstein equation: $\tilR_{\mu\nu}-\frac12\tilg_{\mu\nu}\tilR=\kappa^2S_{\mu\nu}$ as in \cite{white}. }

\subsubsection{Isocurvature Perturbations}

Let us also consider the isocurvature perturbations defined as
\beqa
 \tilS_{ij} = 3 \left( \frac{\delta\tilrho_i}{d\tilrho_i/d\tilt}
                  -\frac{\delta\tilrho_j}{d\tilrho_j/d\tilt}
            \right),
\eeqa
which can be rewritten in terms of the quantities in the Einstein frame
as
\beqa
 \lefteqn{\tilS_{ij} = \frac{3e^a (d\rho_i/dt)(d\rho_j/dt)}
               {\left[(d\rho_i/dt)-4(da/dt)\rho_i\right]
                \left[(d\rho_j/dt)-4(da/dt)\rho_j\right]}
	       } \nonumber \\
  && \qquad        \left[ \frac{\delta \rho_i}{d\rho_i/dt}
                -\frac{\delta\rho_j}{d\rho_j/dt}
                -\frac{4(da/dt)\rho_j}{d\rho_j/dt}
                 \left(\frac{\delta \rho_i}{d\rho_i/dt}
                       -\frac{(d\rho_j/dt)}{(d\rho_i/dt)}
                        \frac{\rho_i}{\rho_j}
                        \frac{\delta\rho_j}{d\rho_j/dt} \right)
                -4\delta a \left(\frac{\rho_i}{d\rho_i/dt}-\frac{\rho_j}{d\rho_j/dt}
                           \right) 
         \right]
\eeqa
with
\beqa
 && \frac{(d\rho_j/dt)}{(d\rho_i/dt)}\frac{\rho_i}{\rho_j} 
   =\frac{da/dt-3H-3(H+da/dt)w_j}{da/dt-3H-3(H+da/dt)w_i}, \\
 && \frac{\rho_i}{d\rho_i/dt}-\frac{\rho_j}{d\rho_j/dt}
   =-3\frac{\rho_i \rho_j}{(d\rho_i/dt)(d\rho_j/dt)} (w_j-w_i)(H+da/dt).
\eeqa
Here, $w_i \equiv p_i/\rho_i$ and we have assumed that the energy
momentum tensor for each component $i$ in the Jordan frame is conserved
individually. It is manifest that $\tilS_{ij} =0$ is equivalent to
$S_{ij} = 0$ if and only if $w_i = w_j$. Thus, the notion of the
adiabaticity is {\it not} conformal-frame independent in general, though
it is again gauge-invariant.

However we believe that this does not immediately imply that the entropy
perturbation is not observable.  Once we fix the frame to define the
unit, it is still meaningful to talk about the entropy perturbation.
The situation is similar to the fact that a statement like 'Newton
constant is changing with time' is meaningless while the statement
'Newton constant in SI units is changing with time' does make sense
\cite{flanagan}.
%Likewise, it is meaningless to talk about the coordinate-dependent quantity, while
% it does make sense to talk about a component in the comoving coordinate system. 

\subsubsection{Relation Among Various Curvature Perturbations}

Here, let us also consider other definitions of curvature perturbations which
are often discussed in the literatures and study the relation among them. 
The equation of
motion  in the Jordan frame (\ref{eom:st:1}) can be rewritten as
\beqa
 \tilR_{\mu\nu}-\frac12 \tilg_{\mu\nu}\tilR = \kappa^2 \tilS_{\mu\nu},
\eeqa
where the "effective" energy momentum tensor $\tilS_{\mu\nu}$ is
defined by
\beqa
 \tilS_{\mu\nu} &=& \frac{\tilT_{\mu\nu}}{\Phi}+
 \frac{1}{\kappa^2 \Phi} \left[ 
   \frac{\omega(\Phi)}{\Phi}
\left(\partial_{\mu}\Phi\partial_{\nu}\Phi-\frac12\tilg_{\mu\nu}(\tilnabla\Phi)^2\right)
+\tilnabla_{\mu}\tilnabla_{\nu}\Phi-\tilg_{\mu\nu}\widetilde{\Box}\Phi-\frac12
 \tilg_{\mu\nu}U \right] \\
 &=& e^{2a} \tilT_{\mu\nu}+
 \frac{1}{\kappa^2} \left[ 
   4\left(\omega(a)+1\right) \partial_{\mu}a\partial_{\nu}a
   -2\tilnabla_{\mu}\tilnabla_{\nu}a
   +\tilg_{\mu\nu} \left( -(2\omega(a)+4)(\tilnabla a)^2
        +2\widetilde{\Box}a-\frac12 U e^{2a} \right)
   \right].
\eeqa
From the Bianchi identity, this "effective" energy momentum tensor is also
covariantly conserved, that is, $\tilnabla_{\mu}{\tilS^{\mu}}_{\nu}=0$. It should be
noticed that the conservation of the original energy momentum tensor
$\tilnabla_{\mu}{\tilT^{\mu}}_{\nu}=0$ in Eq. (\ref{emc:st}) together
with the scalar equation of motion (\ref{eom:st:2}) is equivalent to the
conservation of this new energy momentum tensor.

Then, the following curvature perturbation is often used in the
literatures \cite{white,Kaiser:2010yu},
\beqa
  \tilzeta_s \equiv \tilPhi-\frac{\delta\tilrho_s}{d\tilrho_s/d\tilt}\tilH,
\eeqa
where $\tilrho_s$ and $\delta\tilrho_s$ are the energy density and its
fluctuation defined by the energy momentum tensor $\tilS_{\mu\nu}$
and are given by
\beqa
  \tilrho_s &=& \frac{\tilrho}{\Phi} + \frac{1}{\kappa^2 \Phi} 
                \left[\frac12 \frac{\omega}{\Phi} 
                \left(\frac{d\Phi}{d\tilt}\right)^2 
                - 3\tilH \frac{d\Phi}{d\tilt} + \frac12 U \right] \\
            &=& e^{2a} \tilrho + \frac{1}{\kappa^2}
   \left[ 2\omega \left(\frac{da}{d\tilt}\right)^2
          +6\tilH \frac{da}{d\tilt} + \frac12 U e^{2a} \right], \\
 \delta\tilrho_s &=& \frac{\delta\tilrho}{\Phi}
   + \frac{1}{\kappa^2\Phi}
   \left[\frac{\omega}{2\Phi} 
         \left(2\frac{d\Phi}{d\tilt}\frac{d\delta\Phi}{d\tilt}+
               \left(\frac{d\Phi}{d\tilt}\right)^2
               \left(\frac{d\omega}{d\Phi}\frac{\delta\Phi}{\omega}
                     -\frac{\delta\Phi}{\Phi}-2\tilPsi \right)          
               \right)
         -3\tilH\left(\frac{d\delta\Phi}{d\tilt}+\tilH\delta\Phi\right)
   \right. \nonumber \\
 && \quad \left.
         +3\frac{d\Phi}{d\tilt}\left(2\tilH\tilPsi-\frac{d\tilPhi}{d\tilt}\right) 
         +\frac12\frac{dU}{d\Phi}\delta\Phi + \frac{\nabla^2
  \delta\Phi}{\tilR^2} \right] \\
 &=&  e^{2a} \delta\tilrho
   + \frac{1}{\kappa^2}
   \left[\left( 4 \omega \frac{da}{d\tilt} +6\tilH\right) \frac{d\delta a}{d\tilt} 
        + \left(2 \frac{d\omega}{da} \left(\frac{da}{d\tilt}\right)^2 
                -4 \omega \left(\frac{da}{d\tilt}\right)^2 
                + 6 \tilH^2 - 12 \tilH \frac{da}{d\tilt}
                + \frac12 \frac{dU}{da} e^{2a} \right) \delta a 
 \right. \nonumber \\
 && \quad \left. 
       - \left( 4 \omega \left(\frac{da}{d\tilt}\right)^2 
                +12 \tilH \frac{da}{d\tilt} \right) \tilPsi
        + 6 \frac{da}{d\tilt} \frac{d\tilPhi}{d\tilt}
        - 2 \frac{\nabla^2 \delta a}{\tilR^2} \right],
\eeqa      
where $\tilPhi$ and $\tilPsi$ are the metric perturbations defined in
Eq. (\ref{eq:metricpert}). 

Similarly, the equation of motion in the Einstein frame (\ref{eom:ein:1})
can be rewritten as
\beqa
  R_{\mu\nu}-\frac12 g_{\mu\nu}R=\kappa^2 S_{\mu\nu},
\eeqa
where another energy momentum tensor $S_{\mu\nu}$ in the Einstein frame
is defined by
\beqa
  S_{\mu\nu} &=& T_{\mu\nu}+\partial_{\mu}\vp\partial_{\nu}\vp-\frac12
                 g_{\mu\nu}\left((\nabla\vp)^2+2V(\vp)\right) \\
             &=& T_{\mu\nu}
   +\frac{4\omega(a)+6}{\kappa^2}\partial_{\mu}a\partial_{\nu}a
   -\frac12 g_{\mu\nu}\left( \frac{4\omega(a)+6}{\kappa^2} (\nabla a)^2
     +2V(a) \right).
\eeqa
From the Bianchi identity, this energy momentum tensor is 
covariantly conserved,
that is, $\nabla_{\mu}{S^{\mu}}_{\nu}=0$, though the original energy
momentum tensor is not conserved, $\nabla_{\mu}{T^{\mu}}_{\nu}=\frac{d
a}{d\vp}(\partial_{\nu}\vp) T$, as given in Eq. (\ref{emc:ein}).
However, $\nabla_{\mu}{S^{\mu}}_{\nu}=0$ can be derived by combining
Eq. (\ref{emc:ein}) and the scalar equation of motion (\ref{eom:ein:2}).
 
Associated with this new energy momentum tensor, the following curvature
perturbation is often used in the literatures (again not to be
confused $\Phi$ with the Brans-Dicke scalar field),
\beqa
  \zeta_s \equiv \Phi-\frac{\delta\rho_s}{d\rho_s/dt}H,
\eeqa
where $\rho_s$ and $\delta\rho_s$ are the energy density and its
fluctuation defined by the new energy momentum tensor $S_{\mu\nu}$ and
are given by
\beqa
  \rho_s &=& \rho + \frac{2\omega+3}{\kappa^2} \left(\frac{da}{dt}\right)^2
             +V, \\
  \delta\rho_s &=& \delta\rho + \frac{4\omega+6}{\kappa^2} 
       \left( \frac{da}{dt}\frac{d\delta a}{dt} 
             - \left(\frac{da}{dt}\right)^2 \Psi \right)
       + \frac{2}{\kappa^2} \frac{d\omega}{da}
         \left(\frac{da}{dt}\right)^2 \delta a
       + \frac{dV}{da} \delta a.
\eeqa

First of all, let us consider the relation between the two curvature
perturbations $\zeta$ and $\zeta_s$ in the Einstein frame. The two
curvature perturbations $\zeta$ and $\zeta_s$ are equal if and only if
\beqa
  \left[ \frac{4\omega+6}{\kappa^2} \frac{d^2a}{dt^2}
       + \frac{2}{\kappa^2} \frac{d\omega}{da} \left(\frac{da}{dt}\right)^2
       + \frac{dV}{da} \right] 
  \left( \frac{\delta\rho}{d\rho/dt} - \frac{\delta a}{da/dt} \right)
  - \frac{4\omega+6}{\kappa^2} \frac{1}{(da/dt)}
  \left[ \frac{da}{dt}\frac{d \delta a}{dt}-\frac{d^2a}{dt^2}\delta a
   - \left(\frac{da}{dt}\right)^2 \Psi
  \right] 
 = 0.
%  \frac{\delta\rho}{d\rho/dt} = \frac{\left[((4\omega+6)/\kappa^2) 
%       \left( (da/dt)(d\delta a/dt) 
%             - (da/dt)^2 \Psi \right) / \delta a
%       + (2/\kappa^2) (d\omega/da) (da/dt)^2
%       + (dV/da)\right] \delta a}
%       {\left[((4\omega+6)/\kappa^2) (d^2a/dt^2)
%       + (2/\kappa^2) (d\omega/da) (da/dt)^2
%       + (dV/da) \right] (da/dt)}.
\eeqa
This condition holds true if the following two conditions are satisfied
\beqa
  && \frac{\delta\rho}{d\rho/dt} = \frac{\delta a}{da/dt}, \\
  && \frac{da}{dt}\frac{d\delta a}{dt} - \frac{d^2a}{dt^2} \delta a 
             - \left(\frac{da}{dt}\right)^2 \Psi = 0. 
\eeqa
The latter condition can be recast in terms of the canonical scalar
field $\vp$ into
\beqa
  \frac{d\vp}{dt}\frac{d\delta \vp}{dt} - \frac{d^2\vp}{dt^2} \delta \vp 
             - \left(\frac{d\vp}{dt}\right)^2 \Psi = 0,
\eeqa
which implies that the intrinsic non-adiabatic pressure perturbation of
$\vp$ (or $a$) vanishes \cite{gordon}.
Hence, $\zeta = \zeta_s$ if $\delta\rho/(d\rho/dt)=\delta
a/(da/dt)$ holds (that is, the entropy perturbation between matter and $\vp$ is vanishing)
and if the intrinsic entropy perturbation of $\vp$ is vanishing.

The total non-adiabatic pressure (entropy) perturbations in the Einstein
frame are also defined in two ways:
\beqa
&&  \delta P_{\rm nad} = \frac{dp}{dt} \Gamma = \delta p - c_s^2
  \delta\rho, \\
&&  \delta P_{\rm s,nad} = \frac{dp_s}{dt} \Gamma_s = \delta p_s - c_{s,s}^2
  \delta\rho_s,
\eeqa
where $c_s^2=(dp/dt)/(d\rho/dt)$, $c_{s,s}^2=(dp_s/dt)/(d\rho_s/dt)$, and
\beqa
  p_s &=& \rho + \frac{2\omega+3}{\kappa^2} \left(\frac{da}{dt}\right)^2
             -V, \\
  \delta p_s &=& \delta p + \frac{4\omega+6}{\kappa^2} 
       \left( \frac{da}{dt}\frac{d\delta a}{dt} 
             - \left(\frac{da}{dt}\right)^2 \Psi \right)
       + \frac{2}{\kappa^2} \frac{d\omega}{da}
         \left(\frac{da}{dt}\right)^2 \delta a
       - \frac{dV}{da} \delta a.
\eeqa    
Then, the total non-adiabatic pressure (entropy) perturbations are
related as follows,
\beqa
  \frac{d\rho_s}{dt} \delta P_{\rm s,nad} - \frac{d\rho}{dt} \delta P_{\rm nad} 
   &=& 2\frac{dV}{da}\frac{da}{dt}\frac{d\rho}{dt} 
     \left( \frac{\delta\rho}{d\rho/dt} - \frac{\delta a}{da/dt} \right)
   + \frac{4\omega+6}{\kappa^2} \frac{da}{dt}
      \left( 2\frac{dV}{da} + \frac{\delta\rho - \delta p}{\delta a} \right)
     \left[ \frac{da}{dt}\frac{d \delta a}{dt}-\frac{d^2a}{dt^2}\delta a
   - \left(\frac{da}{dt}\right)^2 \Psi \right] 
\nonumber \\ && \hspace{-3.7cm}
  +\left[ \frac{4\omega+6}{\kappa^2} \frac{1}{\delta a} 
         \left( \frac{da}{dt}\frac{d\delta a}{dt}
          -\left(\frac{da}{dt}\right)^2 \Psi \right)
        +\frac{2}{\kappa^2} \frac{d\omega}{da} \left(\frac{da}{dt}\right)^2
        +\frac{dV}{da} \right]  
  \frac{da}{dt}\frac{d\rho}{dt}
  \left[ \frac{\delta a}{da/dt} - \frac{\delta \rho}{d\rho/dt}  
   - c_s^2 \left( \frac{\delta a}{da/dt} - \frac{\delta p}{dp/dt} \right) 
  \right].
\eeqa
It is manifest that $\delta P_{\rm s,nad} = 0$ and $\delta P_{\rm nad} =
0$ are equivalent if the following two conditions are satisfied
\beqa
  && \frac{\delta\rho}{d\rho/dt} = \frac{\delta a}{da/dt}, \\
  && \frac{da}{dt}\frac{d\delta a}{dt} - \frac{d^2a}{dt^2} \delta a 
             - \left(\frac{da}{dt}\right)^2 \Psi = 0. 
\eeqa

Next, let us consider the condition for $\tilzeta = \tilzeta_s$ in the
Jordan frame. After some calculations, it is easily found that the two
curvature perturbations $\tilzeta$ and $\tilzeta_s$ in the Jordan frame
are equal if and only if
\beqa
&&  6\tilH \left[ \frac{d^2a}{d\tilt^2}-2(\omega+1)\left(\frac{da}{d\tilt}\right)^2
                - \left( \tilH + \frac{d\tilH/d\tilt}{\tilH} \right) \right]
  \left( \frac{\delta a}{da/d\tilt} - \frac{\delta\tilrho}{d\tilrho/d\tilt} \right)
\nonumber \\ && \qquad \qquad
    +2\left( 2\omega+3\frac{\tilH}{da/d\tilt} \right)
    \left[ \frac{da}{d\tilt}\frac{d \delta a}{d\tilt}-\frac{d^2a}{d\tilt^2}\delta a
          - \left(\frac{da}{d\tilt}\right)^2 \Psi \right] 
    - \frac{2}{\tilR^2} \nabla^2 \delta a = 0.
\eeqa
This condition is satisfied on superhorizon scales, where the spatial
derivatives can be neglected, if the following two conditions are satisfied
\beqa
  && \frac{\delta\tilrho}{d\tilrho/d\tilt} = \frac{\delta a}{da/d\tilt}, \\
  && \frac{da}{d\tilt}\frac{d\delta a}{d\tilt} - \frac{d^2a}{d\tilt^2} \delta a 
             - \left(\frac{da}{d\tilt}\right)^2 \tilPsi = 0. 
\eeqa
Hence, $\tilzeta = \tilzeta_s$ on superhorizon scales if 
$\delta\tilrho/(d\rho/d\tilt)=\delta a/(da/d\tilt)$ holds (that is, the
entropy perturbation between matter and $\varphi$ is vanishing) and if the intrinsic
entropy perturbation of $a$ is vanishing. It should be noticed that the
above two conditions are equivalent to
\beqa
  && \frac{\delta\rho}{d\rho/dt} = \frac{\delta a}{da/dt}, \\
  && \frac{da}{dt}\frac{d\delta a}{dt} - \frac{d^2a}{dt^2} \delta a 
             - \left(\frac{da}{dt}\right)^2 \Psi = 0. 
\eeqa
Thus, as long as these conditions,  which imply the vanishing of the entropy perturbation between matter and the Brans-Dicke scalar field and the vanishing of the intrinsic entropy perturbation of the Brans-Dicke scalar,  
are satisfied, all of the four curvature perturbations coincide,
\beqa
  \tilzeta_s = \tilzeta = \zeta = \zeta_s
\eeqa
on superhorizon scales.

Finally, we give explicit expressions for the total non-adiabatic pressure (entropy)
perturbations in the Jordan frame. They are also defined in two ways:
\beqa
&&  \delta \widetilde{P}_{\rm nad} = \frac{d\tilp}{d\tilt} \widetilde{\Gamma} =
\delta \tilp - \tilc_s^2 \delta\tilrho, \\
&&  \delta \widetilde{P}_{\rm s,nad} = \frac{d\tilp_s}{d\tilt} \widetilde{\Gamma}_s
= \delta \tilp_s - \tilc_{s,s}^2 \delta\tilrho_s,
\eeqa
where $\tilc_s^2=(d\tilp/d\tilt)/(d\tilrho/d\tilt)$, $\tilc_{s,s}^2=(d\tilp_s/d\tilt)/(d\tilrho_s/d\tilt)$, and
\beqa
  \tilp_s &=& \frac{\tilp}{\Phi} + \frac{1}{\kappa^2 \Phi} 
                \left[\frac12 \frac{\omega}{\Phi} 
                \left(\frac{d\Phi}{d\tilt}\right)^2 +\frac{d^2\Phi}{d\tilt^2}
                +2 \tilH \frac{d\Phi}{d\tilt} - \frac12 U \right] \\
            &=& e^{2a} \tilp + \frac{1}{\kappa^2}
   \left[ 2(\omega+2) \left(\frac{da}{d\tilt}\right)^2
          -4\tilH \frac{da}{d\tilt} -2\frac{d^2a}{d\tilt^2} 
          - \frac12 U e^{2a} \right], \\
 \delta\tilp_s &=& \frac{1}{\Phi}\left( \delta\tilp - \tilp_s \delta\Phi \right)
   + \frac{1}{\kappa^2\Phi}
   \left[\frac{\omega}{2\Phi} 
         \left(2\frac{d\Phi}{d\tilt}\frac{d\delta\Phi}{d\tilt}+
               \left(\frac{d\Phi}{d\tilt}\right)^2
               \left(\frac{d\omega}{d\Phi}\frac{\delta\Phi}{\omega}
                     -\frac{\delta\Phi}{\Phi}-2\tilPsi \right)          
               \right)
         +\frac{d^2\delta\Phi}{d\tilt^2}+2\tilH \frac{d\delta\Phi}{d\tilt}
   \right. \nonumber \\
 && \quad \left.
         -2\left(\frac{d^2\Phi}{d\tilt^2}+2\tilH\frac{d\Phi}{d\tilt}\right)\tilPsi
         -\frac{d\Phi}{d\tilt}\left(\frac{d\tilPsi}{d\tilt}-2\frac{d\tilPhi}{d\tilt}
                              \right)  
         -\frac12\frac{dU}{d\Phi}\delta\Phi - \frac23\frac{\nabla^2
  \delta\Phi}{\tilR^2} \right] \\
 &=&  e^{2a} \left( \delta\tilp +2\tilp \delta a \right)
   + \frac{1}{\kappa^2}
   \left[ - 2 \frac{d^2\delta a}{d\tilt^2}
        + 4 \left( (\omega+2) \frac{da}{d\tilt} -\tilH\right) \frac{d\delta a}{d\tilt} 
        + \left( 2\frac{d\omega}{da} \left(\frac{da}{d\tilt}\right)^2 
                -U e^{2a} - \frac12 \frac{dU}{da} e^{2a} \right) \delta a 
 \right. \nonumber \\
 && \quad \left. 
       + 4 \left( \frac{d^2a}{d\tilt^2} + 2 \tilH \frac{da}{d\tilt} 
                 - (\omega+2) \left(\frac{da}{d\tilt}\right)^2 
           \right) \tilPsi
        + 2 \frac{da}{d\tilt} \frac{d\tilPsi}{d\tilt}
        - 4\frac{da}{d\tilt} \frac{d\tilPhi}{d\tilt}
        + \frac43 \frac{\nabla^2 \delta a}{\tilR^2} \right],
\eeqa      
where $\tilPhi$ and $\tilPsi$ are the metric perturbations defined in
Eq. (\ref{eq:metricpert}). 
Then, the total non-adiabatic pressure (entropy) perturbations in the
Jordan frame are related as follows,
\beqa
  \frac{d\tilrho_s}{d\tilt} \delta \tilP_{\rm s,nad} 
   - \frac{6}{\kappa^2} \tilH \frac{d\tilH}{d\tilt} e^{2a} \delta
  \tilP_{\rm nad} 
&=&
  \frac{6}{\kappa^2} \tilH \frac{d\tilH}{d\tilt} e^{2a}
   \left( \frac{d\tilp}{d\tilt}-\frac{2}{3\left(\tilH-da/d\tilt \right)} 
          \frac{da}{d\tilt} \frac{\tilrho}{d\tilt} \right)
  \left( \frac{\delta\tilrho}{d\tilrho/d\tilt}-\frac{\delta a}{da/d\tilt} \right) 
\nonumber \\ && \hspace{-5cm}
  + \frac{6}{\kappa^4} \tilH \frac{d\tilH}{d\tilt}
    \left[ \frac43 \frac{\tilH}{\left( \tilH-da/d\tilt \right)}
  \frac{\nabla^2\delta a}{\tilR^2}  
         - \frac{2}{da/d\tilt} \frac{d}{d\tilt}
         \left( \frac{da}{d\tilt}\frac{d\delta a}{d\tilt}
          -\left(\frac{da}{d\tilt}\right)^2 \tilPsi -
                 \frac{d^2a}{d\tilt^2}\delta a \right)
    \right.
\nonumber \\ && \hspace{-3cm} \left.
        + \left\{ \frac{4}{\tilH-da/d\tilt} \left(
  \omega\left(\tilH-\frac53\frac{da}{d\tilt} \right) -
  \frac{\tilH^2}{da/d\tilt} \right) +8  \right\}  
         \left( \frac{da}{d\tilt}\frac{d\delta a}{d\tilt}
           -\left(\frac{da}{d\tilt}\right)^2 \tilPsi -
                 \frac{d^2a}{d\tilt^2}\delta a \right)
        \right] 
\nonumber \\ && \hspace{-5cm} 
  - \frac{d\tilp_s}{d\tilt} \frac{\tilH}{\tilH-da/d\tilt}
    \left[ e^{2a} \frac{d\tilrho}{d\tilt} 
      \left( \frac{\delta\tilrho}{d\tilrho/d\tilt}-\frac{\delta a}{da/d\tilt} \right) 
          +\frac{2}{\kappa^2} \left\{
            \left(2\omega+3\frac{\tilH}{da/d\tilt}\right)
            \left( \frac{da}{d\tilt}\frac{d\delta a}{d\tilt}
          -\left(\frac{da}{d\tilt}\right)^2 \tilPsi -
                 \frac{d^2a}{d\tilt^2}\delta a \right)   
          -\frac{\nabla^2\delta a}{\tilR^2} \right\}
    \right]. 
\eeqa
Here we have used the following relations,
\beqa
  \frac{d\tilrho_s}{d\tilt} &=& \frac{6}{\kappa^2} \tilH
  \frac{d\tilH}{d\tilt}, \\ 
  \delta \tilrho_{s} &=& \frac{6}{\kappa^2} \tilH \frac{d\tilH}{d\tilt}
  \frac{\delta a}{da/d\tilt} + \frac{1}{\kappa^2} \frac{\tilH}{\tilH-da/d\tilt}
   \left[ \kappa^2 e^{2a} \left(\delta\tilrho-\frac{d\tilrho/d\tilt}{da/d\tilt}
  \delta a \right) \nonumber \right. \\ &&
  \left. \qquad \qquad
  + 2\left(2\omega+3\frac{\tilH}{da/d\tilt}\right)
      \left( \frac{da}{d\tilt}\frac{d\delta a}{d\tilt}
          -\left(\frac{da}{d\tilt}\right)^2 \tilPsi -
                 \frac{d^2a}{d\tilt^2}\delta a \right)  
      -2\frac{\nabla^2\delta a}{\tilR^2} 
      \right], \\
  \frac{d\tilPhi}{d\tilt} &=& \tilH\tilPsi
    +\frac{d\tilH/d\tilt}{da/d\tilt} \delta a
   + \frac{1}{6} \frac{1}{\tilH-da/d\tilt}
   \left[ \kappa^2 e^{2a} \left(\delta\tilrho-\frac{d\tilrho/d\tilt}{da/d\tilt}
  \delta a \right)  \nonumber \right. \\ &&
  \left. \qquad \qquad
    + 2\left(2\omega+3\frac{\tilH}{da/d\tilt}\right)
      \left( \frac{da}{d\tilt}\frac{d\delta a}{d\tilt}
          -\left(\frac{da}{d\tilt}\right)^2 \tilPsi -
                 \frac{d^2a}{d\tilt^2}\delta a \right)  
      -2\frac{\nabla^2\delta a}{\tilR^2} 
      \right].
\eeqa     
It is manifest that $\delta \tilP_{\rm s,nad} = 0$ and $\delta
\tilP_{\rm nad} = 0$ are equivalent on superhorizon scales if the
following two (three) conditions are satisfied
\beqa
  && \frac{\delta\tilrho}{d\tilrho/d\tilt} = \frac{\delta a}{da/d\tilt}, \\
  && \frac{da}{d\tilt}\frac{d\delta a}{d\tilt} - \frac{d^2a}{d\tilt^2} \delta a 
             - \left(\frac{da}{d\tilt}\right)^2 \tilPsi = 0, \\ 
  && \frac{d}{d\tilt} \left( 
      \frac{da}{d\tilt}\frac{d\delta a}{d\tilt} - \frac{d^2a}{d\tilt^2} \delta a 
             - \left(\frac{da}{d\tilt}\right)^2 \tilPsi \right) = 0.
\eeqa

\section{Summary}

We have discussed the frame independence/dependence of the observable
quantities between the Jordan and the Einstein frame in scalar-tensor
gravity and have provided a Jordan-Einstein dictionary. 
We have found that dimensionless quantities and the relations of them (redshift, magnitude redshift relation, reciprocity relation, the Sachs-Wolfe effect) are 
independent of the conformal frame. 
The dimensional quantities such as the Hubble parameter and 
the luminosity/angular diameter  distances, are related 
to each frame by a simple change of local units due 
to the conformal transformation, although for the effective gravitational constant the transformation  involves the asymptotic value of the conformal factor. 

As applications, we have considered the horizon problem and the dark energy equation of state. We have found that the condition for solving the horizon 
problem itself is conformal-frame independent. However, the condition that the comoving Hubble distance decreases
 differs in each frame and hence the detailed implementation of the mechanism to solve the horizon problem can
 differ depending on the frame.  
We have also considered the situation where an observer (erroneously) interprets the observational 
data using the Einstein gravity with a minimal coupling to matter although 
the true gravity theory is scalar-tensor gravity and have found that   
 the apparent equation of state dark energy 
could be "phantom-like" $w<-1$.

Finally we have studied the frame dependence of 
 curvature perturbations. 
We have found that the absence of the non-adiabatic pressure and hence 
the absence of the isocurvature 
 perturbations are conformal-frame dependent in general. 
There exist four types of curvature perturbations on uniform-matter density hypersurfaces depending on the definition of the (effective) energy momentum 
tensors $\tilT_{\mu\nu}$ (or $T_{\mu\nu}$) and $\tilS_{\mu\nu}$ (or $S_{\mu\nu}$). 
For the matter defined by the energy momentum tensors $\tilT_{\mu\nu}$ 
and $T_{\mu\nu}$, the conformal invariance of the corresponding curvature perturbations on uniform-density hypersurfaces  $\tilzeta$ and $\zeta$ holds if 
the entropy perturbation between matter and the Brans-Dicke scalar field is vanishing or the equation of state of matter is constant. 
For the matter defined by the energy momentum tensors $\tilS_{\mu\nu}$ 
and $S_{\mu\nu}$, the corresponding curvature perturbations on uniform-density hypersurfaces  $\tilzeta_s$ and $\zeta_s$ coincide with $\tilzeta$ and $\zeta$ 
if the entropy perturbation between matter and the Brans-Dicke scalar field is vanishing and if the intrinsic entropy perturbation of the Brans-Dicke 
scalar is vanishing.

Although our analysis in this paper was limited to classical theory, 
we would like to extend the analysis to quantum theory (see \cite{np} 
for recent attempts) in order to investigate  
to what extent the conformal-frame independence persists in quantum theory \cite{cnsy}.

%%%%%%%%%%%%%%%%%%%%%%%%%%%%%%%%%%%%%%%%%%%%%%%%%%%%%%%%%%%%%%%%%%%%%%
\section*{Acknowledgments}
%%%%%%%%%%%%%%%%%%%%%%%%%%%%%%%%%%%%%%%%%%%%%%%%%%%%%%%%%%%%%%%%%%%%%%
We would like to thank Misao Sasaki for useful discussion. 
This work was supported in part by a Grant-in-Aid for Scientific
Research from JSPS (Nos.\,24540287(TC), 24111706(MY), and 25287054(MY))
and in part by Nihon University (TC).

%
%%%%%%%%%%%%%%%%%%%%%%%%%%%%%%%%%%%%%%%%%%%%%%
\appendix
\section{The equations of motion in FRW universe}
\label{app1}
%%%%%%%%%%%%%%%%%%%%%%%%%%%%%%%%%%%%%%%%%%%%%%
In this appendix, we list the equations of motion for FRW universe models for completeness. 

The equations of motion for FRW universe models in the Jordan frame are given by
\beqa
&&3\Phi\left(\tilH^2+\frac{K}{\tilR^2}\right)={\kappa^2\tilrho}+\frac{\omega}{2\Phi}\left(\frac{d\Phi}{d\tilt}\right)^2-3\tilH\frac{d\Phi}{d\tilt}+\frac12 U(\Phi),\\
&&\Phi\left(\frac{d\tilH}{d\tilt}-\frac{K}{\tilR^2}\right)=-\frac{\kappa^2}{2}(\tilrho+\tilp)-\frac12\frac{d^2\Phi}{d\tilt^2}+\frac12\tilH\frac{d\Phi}{d\tilt}-\frac{\omega}{2\Phi}\left(\frac{d\Phi}{d\tilt}\right)^2
,\\
&&\frac{d^2\Phi}{d\tilt^2}+3\tilH\frac{d\Phi}{d\tilt}=\frac{1}{2\omega+3}\left(
\kappa^2(\tilrho-3\tilp)-\frac{d\omega}{d\Phi}\left(\frac{d\Phi}{d\tilt}\right)^2-\Phi\frac{dU}{d\Phi}+2U\right),
\eeqa
and the energy-momentum conservation is given by
\beqa
\frac{d\tilrho}{d\tilt}+3\tilH(\tilrho+\tilp)=0.
\label{tilrho_m}
\eeqa
The equations of motion in the Einstein frame are given by
\beqa
&&3\left(H^2+\frac{K}{R^2}\right)=\kappa^2\left(\rho+
\frac12\left(\frac{d\vp}{dt}\right)^2+V \right)\equiv \kappa^2(\rho+\rho_{\vp}),\label{friedmann:e}\\
&&\frac{dH}{dt}-\frac{K}{R^2}=-\frac{\kappa^2}{2}\left(\rho+p+\left(\frac{d\vp}{dt}\right)^2\right),\\
&&\frac{d^2\vp}{dt^2}+3H\frac{d\vp}{dt}+\frac{dV}{d\vp}=\frac{da}{d\vp}(-\rho+3p),
\eeqa
and the energy-momentum conservation is given by
\beqa
\frac{d\rho}{dt}+3H(\rho+p)=-\frac{d\vp}{dt}\frac{da}{d\vp}(-\rho+3p).
\label{em-conserv:e}
\eeqa
Considering the correspondence between the variables in the Jordan frame and those in the Einstein frame given by (see also Eq. (\ref{Phi-vp}), Eq. (\ref{V-U}) and Eq. (\ref{tau:scale}))
\beqa
{\tilH=e^{-a}H_E=e^{-a}\left(H+\frac{da}{dt}\right)},~~~~\tilrho=e^{-4a}\rho,~~~~\tilp=e^{-4a}p,~~~~\Phi=e^{-2a},
\eeqa
we verify that the equations of motion in the both frames are
equivalent. {Of course, this is a special case of the equivalence of
the equations of motion in the both frames, as mentioned in the section
II.}

%%%%%%%%%%%%%%%%%%%%%%%%%%%%%%%%%%%%%%%%%%%%%%
%\section{The evolution equation of and the conservation of $\zeta$}
%\label{app2}
%%%%%%%%%%%%%%%%%%%%%%%%%%%%%%%%%%%%%%%%%%%%%%
%In this appendix, we derive the evolution equation of $\zeta$ from 
%the divergence of the energy-momentum tensor, $\nabla_{\mu}{T^{\mu}}_{\nu}=\frac{d
%a}{d\vp}(\partial_{\nu}\vp) T$, and derive the condition  for the conservation of $\zeta$. 

%%%%%%%%%%%%%%%%%%%%%%%%%%%%%%%%%%%%%%%%%%%%%%%%%%%

%%%%%%%%% references %%%%%%%%%%%%%%%%%%%%%%%%%%%%%%

\end{document}